# A simple method for the characterization of HPGe detectors

Patrice Medina, Cayetano Santos, Denis Villaumé
IReS, IN2P3 – CNRS / Université Louis Pasteur
23 rue du Loess, BP28, Strasbourg Cedex 2, F67037 - France

*Abstract – A functional approach to the concept of pulse shape synthesis is introduced here. This paper reviews some of the mechanisms influencing the process of signal induction, and the solution of the electric field and charge carriers drift phenomena are considered for arbitrary crystal geometries. Then, performing some well-known principles and a set of simple techniques provides an open way to the implementation of pulse shape analysis algorithms.*

*Keywords – simulation, segmented HPGe detectors, pulse shape, γ ray tracking, anisotropy.*

## I. INTRODUCTION

Germanium detectors have been in use for more than 30 years. However, new capabilities are still being developed and constant improvements are made, such as the use of high-purity (HPGe) materials with a residual impurity content of down to less than $10^{10}$ $cm^{-3}$ to improve energy resolution, or the development of highly segmented detectors with low noise and fast preamplifiers. Recent techniques [1, 2] enabling the tracking of γ radiation will contribute greatly to improve performances of high-spatial resolution detectors.

Based on the segmentation of the outer contacts and digital pulse shape analysis, emerging advances will enable to locate γ ray interaction position. The concept of γ tracking will perform Compton suppression and Doppler correction, determining the time sequence of interactions with high accuracy and enabling the characterization of unknown γ ray sources. Such a detector will be characterized by a high efficiency and a good peak-to-background ratio. The two tracking arrays presently developed are the European AGATA [3] and the American GRETA [4] projects. Their geometries are based on closed-end coaxial tapered crystals shaped with a hexagonal cross-section (Fig. 1).

The location of gamma-ray interactions will be enabled by the use of digital electronics and on-line pulse shape analysis algorithms. In this context, pulse shape analysis is a key issue. It involves an in-depth understanding on how the induction process takes place, as well as on which are the processes providing contact signals. Computer aided tools suppose a powerful instrument to this goal. How custom numerical simulation can help to understand aspects of this field of research is our main topic of discussion. This paper reviews some basic concepts related to pulse shape generation [5, 6], and reports on some typical examples whose use has spread since last years.

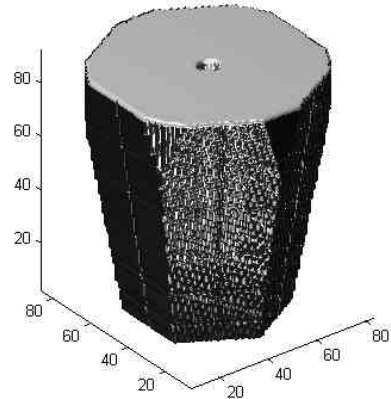

Fig. 1 AGATA crystal geometry.

## II. PROPOSED APPROACH

The aim of this paper is to present a comprehensive methodology to the characterization of HPGe detectors by numerical methods, with no restriction to a particular geometry. The *multi geometry simulation* (MGS) code is structured in a progressive way, starting from the definition of the crystal geometry to characterize, up to the generation of the expected pulse shapes at the contacts. The global process is split into several stages:

- Calculation of the electric field starting from the solution of Poisson equation

$$\nabla^2 \varphi = -\rho / \varepsilon \qquad \text{Eq. 1}$$

- Implementation of charge carrier transport in a semi-conducting medium [7]
- Trajectories of charge carriers for arbitrary interaction points
- Application of the Ramo [5, 6] theorem providing the resulting charge recovery at the contacts
- Weighting potential and weighting field [5, 6] resolution
- Scanning of selected areas in the crystal
- Simulation of the charge-collection efficiency.

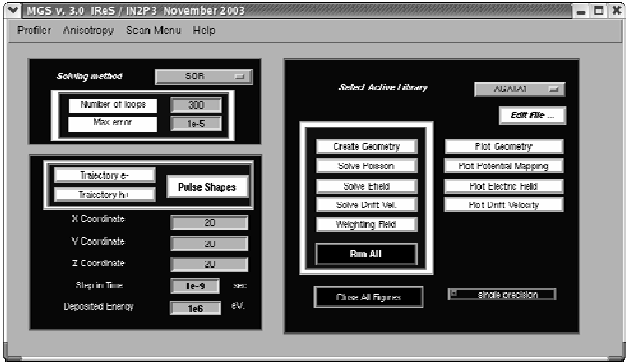

Fig. 2 Main interface to the libraries under Linux Red Hat 9.

To facilitate the characterization of new crystal geometries, a set of customisable templates representing standard, well-known detectors is included. Finally, all routines are accessible through a graphical user interface (Fig. 2) allowing interactive tuning of parameters (time step, deposited energy, interaction position, etc.), along with dedicated plotting subroutines for display. The whole package is available as a stand-alone graphical application running under Windows / Linux PC.

MatLab [8] is the selected environment to develop our algorithms. It is a matrix-based simulation language, as well as an environment intended for numerical and symbolic computations and graphics. It is built on a matrix-oriented programming language especially designed for fast matrix computations, suitable for grid-based solving algorithms. This enables the modelling of general geometries with the support of a cubic grid, whose pitch can be adapted as a function of the available computer resources and the desired accuracy. The expected modelling capabilities are limited by the computing power, the available memory storage capacity and the grid size. We use interpolation to get finer details when necessary.

The computing techniques described here intend to take advantage of the above possibilities to fully characterize arbitrary detectors in a straightforward way. This will be possible thanks to several well-known numerical algorithms [9, 10], and the simplicity of their implementation using the matrix-oriented language.

We emphasize on the fact that dimensions are arbitrary, as long as the ratio detector dimensions to grid step imposes a limit to data storage and management. For example, we characterize the AGATA geometry using a one-millimeter grid for a nine centimeters length detector, which accounts for a one million-point matrix. For a one centimeter detecting volume, a finer mesh (~100 μm) could be used making use of the same computing resources.

## A. Solution of Poisson's equation

A single "static" function $u$ $(x, y, z)$, satisfying the Poisson's equation within some $(x, y, z)$ region of interest needs to be found. In addition, some desired behaviour on the boundary of this region (boundary value problem) must be respected. More exactly, the bias voltage is set to a constant value at the contacts, leading to a Dirichlet boundary problem. We will assume in the following that a grounded, vacuum chamber encloses the detecting volume and that an intrinsic space charge density $\rho = e\, N_{A/D}$, exist where e is the elementary charge, and $N_{A/D}$ is the density of acceptors or donors of a p- or n-type detector. A gradient in space charge density along the crystal is considered.

The principal algorithms used here to solve the Poisson's equation are the so-called *direct, relaxation* [9] and *SOR* [10] methods. Due to the requisite of satisfying all boundary conditions simultaneously, this problem comes to the solution of a large number of simultaneous linear algebraic equations (*direct* method). Alternatively, they can be solved by linear approximation and successive iterations (Jacobi's or *relaxation* method). Further improvements can be achieved by more elaborate techniques, leading to the *SOR* method. As stability is relatively easy to achieve by the methods used here, the efficiency of the algorithm becomes the main concern.

In this way, we can consider the solution of equation 1 by the *finite difference method*, which represents the function $u$ $(x, y, z)$ by its values at the discrete set of points characterized by a grid spacing $h$. Substituting the equation by its finite-difference representation at mesh points provides a system of linear equations.

In order to write this system in matrix form, we consider these points as a vector, by numbering the three dimensions of the grid points in a single one-dimensional sequence of values. We define a new index, $n = i + (j-1)\, I + IJ\, (k-1)$, where i, j and k are the indexes of the three dimensional mesh points. This leads to the linear system:

$$u(n{+}1) + u(n{-}1) + u(n{+}I) + u(n{-}I) + u(n{+}IJ) + u(n{-}IJ) - 6u(n) = h^2\rho, \quad n = 1\ldots N, \ N{=}I.J.K \qquad \text{Eq. 2}$$

A given subset of points is considered as boundary points where $u$ $(x, y, z)$ has been specified. Now the problem takes the "classical" form

$$A(NxN) * u(1xN) = b(Nx1) \qquad \text{Eq. 3}$$

*Direct methods* [9] to the solution of this equation are more generally applicable if there is enough storage available to carry them out. They will depend either on the capacity to factorise the former system of equations, or on knowing its corresponding eigenvalues and eigenvectors. When the matrix representing this system is positive definite, there is a basis of orthogonal eigenvectors, and it can be found explicitly. Then, the solution of the problem can be expressed as

$$u_N = \sum_{1}^{N} \frac{1}{\lambda_j} \left( b, v^{(j)} \right) * v^{(j)}$$

Eq. 4

where $v^{(j)}$ is the mentioned basis of dimension N corresponding to the eigenvectors $\lambda_j$. Since arbitrary border conditions are expected to occur, the matrix will not always be symmetric, and this approach will not be possible at low computing costs.

The second strategy is the *iteration method* [9], where an initial guess is chosen and then we solve by iterating until a solution is found. The algorithm consists on using the average of $u$ $(x, y, z)$ at its six nearest-neighbour points on the grid plus the contribution from the source, representing the density of impurities. The procedure is then iterated until convergence is reached when a worst-case checking of the difference between previous and current values is fulfilled. This approach involves multiplication by a constant and addition of shifted matrices, yielding it ideal for an implementation in MatLab language, both in computational and load requirements, as well as in coding efficiency.

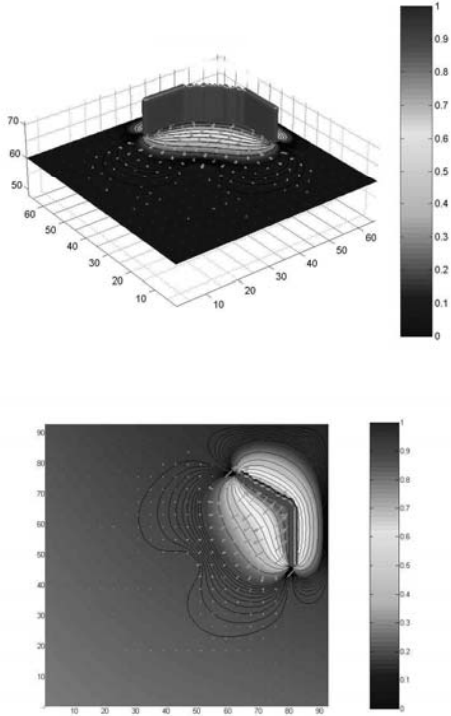

Fig. 3 Weighting potential mapping for one of the segments of an TIGRE (top) and AGATA detector (bottom).

More elaborated methods, such as *successive over relaxation* (SOR) [10] provide faster convergence, making an over correction to the value of the *r*th stage of previous iterations, thus anticipating future corrections. Optimal SOR requires of the order of $J$ iterations, as opposed to of the order of $J^2$ in *relaxation*, so the number of iterations is proportional, in some functional way, to the number of mesh points. For comparison purposes, relaxation method can take about three hours to solve a one million points matrix, whereas using SOR gives a solution in about ten minutes (Pentium III PC). All three solving algorithms have been implemented.

B. *Drift velocities.*

The experimental evidence for the dependence of pulse shapes in closed-end HPGe detectors on the electron drift velocity anisotropy has been clearly established [7], as well as its influence on the signal processing methods related to pulse shape digitisation, which is required to perform the tracking algorithms. Anisotropy, regarding charge carriers, is a major matter of concern when dealing with semi-conducting devices operating at high-electric fields, where deviation from low-field ohmic behavior is observed.

Electric conduction is to be influenced by several parameters, such as the strength of the applied electric field, the crystal orientation and the lattice temperature. Drift velocity magnitude and angle shift with respect to the electric field direction will be affected by the fact that the tensor of differential mobility will become non-diagonal and field dependent, with decreasing diagonal elements as the strength of the electric field increases.

More generally, in the present work the following expression has been considered when referring to the dependence of the drift velocity of charge carriers on the applied electric field

$$v_d = A(\|E\|) \sum_j \frac{n_j}{n} \frac{\gamma_j * E_0}{(E_0 * \gamma_j * E_0)^{1/2}}$$

Eq. 5

The interpretation of each term can be found in Ref. [7]. Anisotropy can be typically observed when performing a full detector (x, y) scan. In the present work data from the TIGRE detector [11] scanning performed at the Liverpool U. by M. Deskovitch et al. [12] is used. The scanning of the detector front face has been performed by meshes of 2 mm. The data is in good agreement with simulation calculations (fig 4).

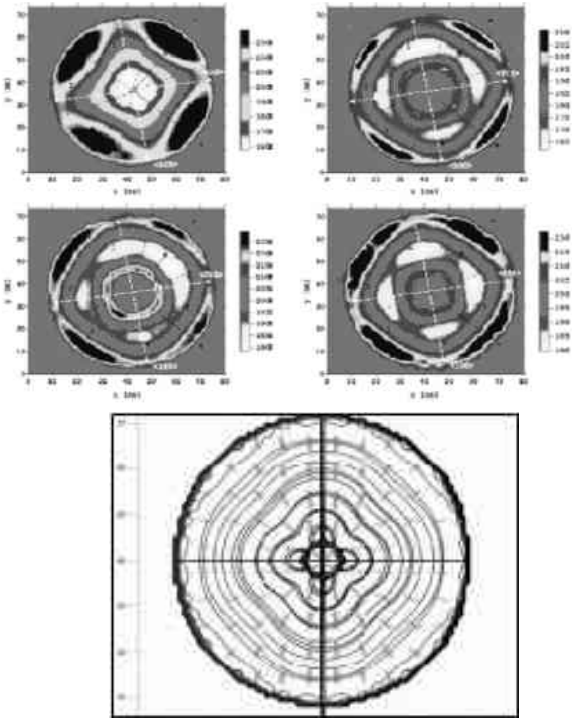

Fig. 4 Experimental photopeak efficiency maps [9] at four different depths along the crystal axis (top panel) face to simulation calculations (bottom panel). The isocontours are in good agreement with experiment.

## C. *Application of Ramo's theorem*

When an incoming γ ray interacts within a segmented detector, it generates free moving charges (electron-hole pairs). The movement of these charges induces a signal in all segments during the carriers drift time (in this work we assume punctual charges). The deposited energy can then be obtained from the output pulse amplitude analysis. Neighboring segments recover a transient signal, whereas in the touched segment a net charge can be measured. Ramo's theorem **[5, 6]** provides the induced current with time as the dot product of the drift velocity and a fictitious entity, the weighting field (Fig. 3), independently of the bias voltage and space charge distribution. Its calculation comes to solving Laplace's equation (Poisson's equation without any sources) with particular border conditions.

Thus, at this stage we take advantage of previous results (drift velocities calculation and potential mapping solution). The efficiency of implementation of this theorem is a key issue. This is due on one hand to the processing power required to calculate charge carrier trajectories, and on the other hand to the need to synthesize in matrix language the induced pulse shapes. Figure 5 provides an example of real (central segment) and induced (surrounding segments) currents (top panel), and charge (button panel) signals calculated for a planar crystal segmented in 3 by 3 pixels.

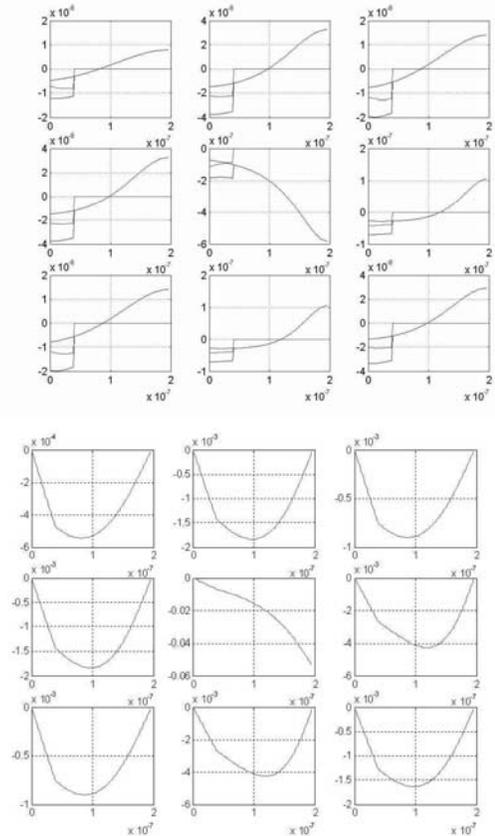

Fig. 5 Induced pulse shapes in a 3 by 3 segmented planar crystal: current induced by holes and electrons (top panel), and charge (bottom panel).

Induced charge is calculated using a model for the charge sensitive preamplifier output stage, simulated as a two-stage transfer function.

## III. SOME RESULTS AND FUTURE WORK

Preliminary, qualitative results were obtained by scanning a 3 by 3-planar detector. Poor collection areas were predicted by simulation (Fig. 6), and confirmed experimentally. In particular, charge collection efficiency was measured by simulation performing a full one-millimeter step scan of the crystal. The resulting efficiency map is shown in Fig. 6 (bottom). The observed pulse shapes confirmed also what was predicted (Fig. 5). The expected behavior is well reproduced by the experimental observation. Simulations helped to develop at IReS a concept allowing the straightening of the electric field, actually under study in the form of a prototype.

Several other typical geometries have been implemented, and are available in the form of customizable templates. Between them, microstrip and CdTe planar detectors, EUROBALL and EXOGAM clover crystals and EUROBALL cluster capsule.

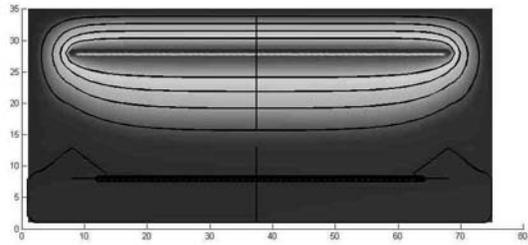

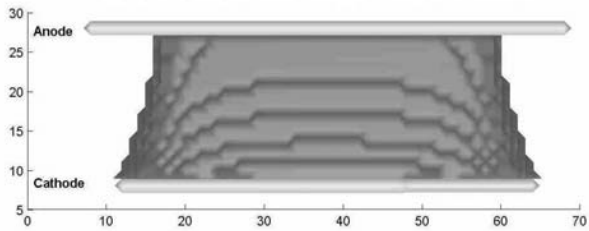

Fig. 6 Potential mapping of a planar detector (top). The upper contact is the anode on which HV is applied. The bottom contact is the segmented cathode which size is reduced compared to the anode by the guard ring. Poor collection areas are expected at the edges due to non-flat equipotential lines and to weak electric fields near the cathode corners. Volume efficacy (bottom) shows areas appropriate to detection.

At present, we concentrate on the comparison with experimental scanned data from the U. of Liverpool. We plan to reproduce the experimental database of a six by six coaxial detector in order to validate quantitatively this work on an experimental basis. The final target is to fully characterize the new AGATA asymmetric capsule before its delivery date.

In parallel, full simulation scanning of the AGATA geometry prototype is being accomplished, in order to develop fast and as simple as possible new pulse-shape methods implementable in FPGA's for on-line calculation. Further improvements to the program, like the multi grid method **[9]** currently under study, should provide a key step to this goal.

## IV. CONCLUSIONS

A simple methodology for the characterization of γ ray detectors is presented here. Electric potential, electric and weighting fields and carrier transport phenomena were considered in order to simulate the induced pulse shapes. This method can be extended to any semi-conducting detector, with no restriction of shape or detecting media, provided the physics of the pulse shape generation are known. For example, in CdTe materials the loss of charge carriers during the drift time cannot be ignored **[13]**, such as is the case in HPGe detectors. In the same way, transport phenomena can be generalized, spreading the presented approach to other non semi-conducting based detectors.

## ACKNOWLEDGMENT

The authors wish to thank Dr. G. Duchêne for helpful discussions on detectors, and for corrections to this manuscript. The authors thank Dr. D. Husson for useful teaching.